\documentclass[11pt]{article}

\usepackage[margin=1.18in,top=1cm]{geometry} 
\usepackage{authblk} 
\usepackage{graphicx} 
\usepackage{amsmath} 
\usepackage[hidelinks]{hyperref} 
\usepackage{array}
\usepackage[numbers]{natbib} 
\usepackage{float}

\usepackage{times} 
\usepackage{helvet} 

\usepackage{setspace}
\usepackage[labelfont=bf,font=small]{caption} 

\usepackage[top=1cm]{geometry}  

\title{Interleaved dual-species arrays of single atoms using a passive optical element and one trapping laser}

\author[1]{\mbox{Chengyu Fang}}
\author[3]{\mbox{Jared Miles}}
\author[3]{\mbox{Jonathan Goldwin}}
\author[3]{\mbox{Martin Lichtman}}
\author[3]{\mbox{Matthew Gillette}}
\author[3]{\mbox{Michael Bergdolt}}
\author[1]{\mbox{Sanket Deshpande}}
\author[2]{\mbox{Sam A. Norrell}}
\author[2]{\mbox{Preston Huft\textsuperscript{\dag}}}
\author[1,2]{\mbox{Mikhail A. Kats\thanks{Corresponding author: \texttt{mkats@wisc.edu}}}}
\author[2,3]{\mbox{Mark Saffman}}

\affil[1]{Department of Electrical and Computer Engineering, University of Wisconsin-Madison, 1415 Engineering Drive, Madison, WI, 53706, USA}
\affil[2]{Department of Physics, University of Wisconsin-Madison, 1150 University Avenue, Madison, WI, 53706, USA}
\affil[3]{Infleqtion, Inc., Madison, WI, 53703, USA}

\date{\vspace{-10ex}} 

\begin{document}

\maketitle

\begingroup
\renewcommand\thefootnote{\textsuperscript{\dag}}
\footnotetext{Present address: QuEra Computing, Inc. 1380 Soldiers Field Road, Boston, MA 02135}
\endgroup


\section*{Abstract}
We demonstrate trapping of individual rubidium (Rb) and cesium (Cs) atoms in an interleaved array of bright tweezers and dark bottle-beam traps, using a microfabricated optical element illuminated by a single laser beam and a 4F system with spatial filtering. Our approach exploits the opposite-sign dynamic polarizabilities of Rb and Cs, ensuring each species is exclusively trapped in either bright or dark sites. The passive optical mask creates optimal trap depths for both species using three transmittance levels while minimizing the optical phase difference, implemented using a variable-thickness absorbing layer of amorphous germanium. This trapping architecture achieves atom loading rates close to 50\% while reducing system complexity compared to conventional methods using active optoelectronic components and/or multiple laser wavelengths.

\section{Introduction}
\noindent Arrays of trapped neutral atoms are being rapidly developed for  quantum sensing, communication, and computing applications\cite{Kaufman2021}.  Scaling up to large numbers of trapped atoms in well-ordered arrays is necessary for large-scale computation and other quantum-information applications. Large arrays have been achieved using sophisticated optical systems comprising high-power lasers and active beam-shaping components such as acousto-optic deflectors (AODs) \cite{Graham2022}, spatial light modulators (SLMs) \cite{Kim2019,Manetsch2024}, and digital mirror devices (DMDs) \cite{Stuart2018,Wang2020}. Recent developments utilize interleaved arrays of two different atomic species for error correction and noise reduction\cite{Beterov2015,Singh2022,Singh2023,Petrosyan2024} where, for example, one type of atom can be a computation qubit and the other a measurement qubit. Trapping multiple species of atoms simultaneously adds to system  complexity, especially when using multiple trapping lasers with different wavelengths and active electro-optical components \cite{Singh2022}.  System complexity can be reduced by using specially fabricated  passive components to generate arrays of 
optical traps\cite{Huft2022,Schlosser2023,RHuang2024,Huang2023,Holman2024,Hsu2022}.

\begin{figure}[H] 
    \centering
    \includegraphics[trim=16pt 0pt 0pt 0pt,clip,width=0.8\linewidth]{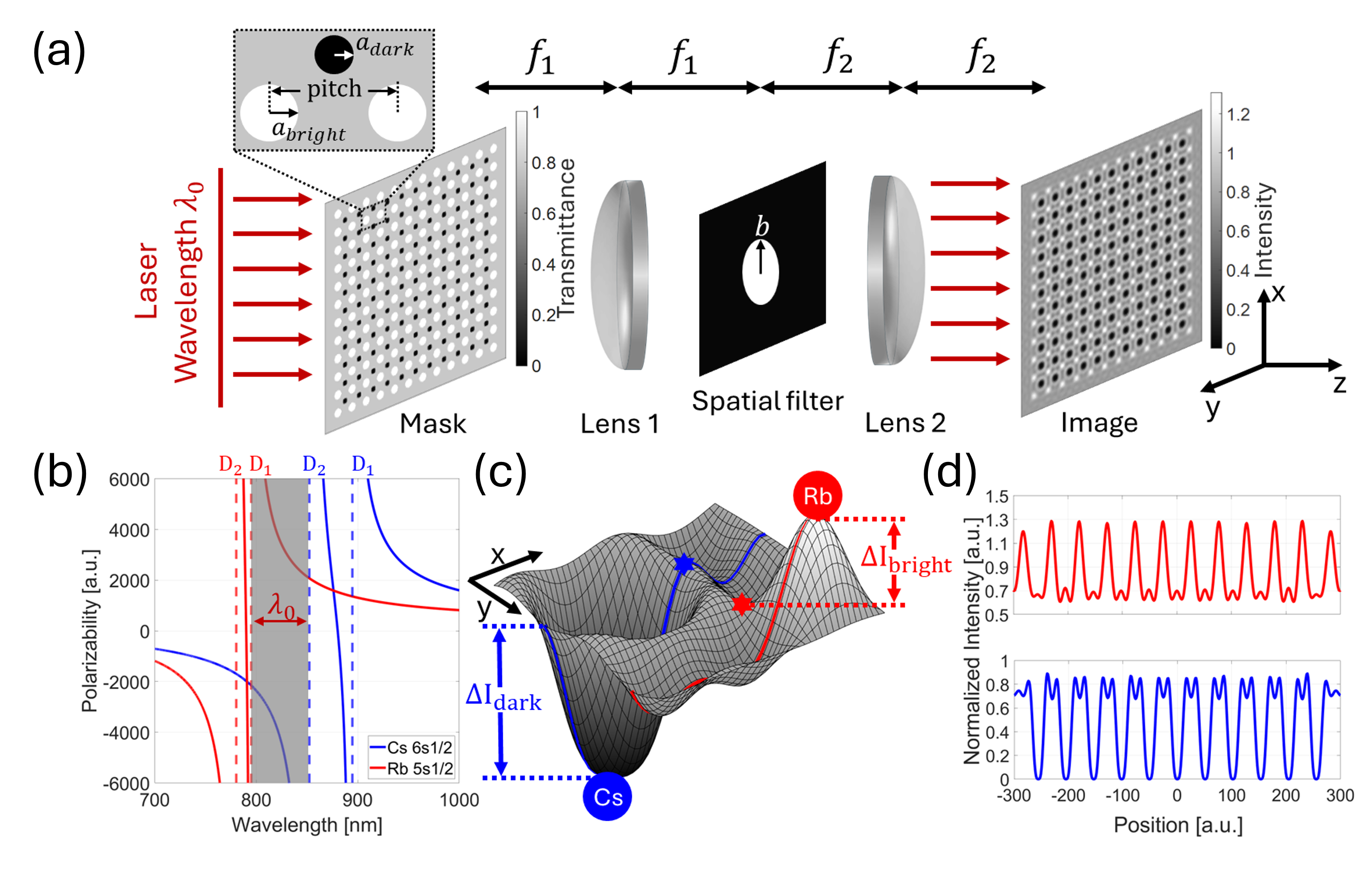}
    \caption{Working principle of the dual-species array generator. \textbf{(a)} Schematic of the 4$f$ system to produce an interleaved array of bright and dark traps. A laser at $\lambda_0$ illuminates the optical mask, followed by a 4$f$ system with a spatial filter, to produce the trapping light at the back focal plane of lens 2. A spatial filter with radius $b$ is placed between lens 1 and lens 2 to filter out high-frequency components to produce a Gaussian-like trapping profile. \textbf{(b)} The wavelength of the input beam $\lambda_0$ should be selected between the Rb D1 line and Cs D2 line, such that the polarizability of one species is positive and the other negative. \textbf{(c)} Calculated intensity contour (vs. position) for one period in the image plane, showing one bright trap for Rb and one dark trap for Cs. The intensity difference $\Delta I$ between the trap centers and the locations identified by blue and red stars represent the intensity depths. \textbf{(d)} The intensity line profiles of dark traps (blue line) and bright traps (red line), as identified in Fig. 1(c). The dark traps and the bright traps are optimized to achieve zero intensity and maximum intensity, respectively, at the trap centers, using a shared spatial filter. The slight intensity variation at the boundary of the trap array is due to the loss of periodicity at the boundary.}
    \label{fig:method_and_design}
\end{figure}

Here, we demonstrate a simple, all-passive optical system that can transform a single high-power laser beam into an array of interleaved dark (blue-detuned) and bright (red-detuned) traps, with each type of trap exclusively hosting a certain species of atom [Fig. \ref{fig:method_and_design}(a, b)]. This system is enabled by a passive optical intensity mask which can be inexpensively fabricated with conventional scalable microelectronics processes (photolithography, thin-film evaporation, liftoff, and wet etching), and generates three levels of transmittance (zero, medium, high) while managing the phase between the different regions, which forms the traps. The mask is highly robust, functioning over a wide range of laser wavelengths with substantial tolerance to fabrication errors. Using this technique with a single 825 nm laser, we experimentally demonstrated interleaved single-atom arrays of $^{87}$Rb in bright traps and $^{133}$Cs in dark traps, each with a loading rate close to 50\%.

\section{Working principle and design}

Our approach, shown in Fig. \ref{fig:method_and_design}(a), is to send a laser beam with single wavelength $\lambda_0$ through a mask with three transmittance levels: fully transmitting regions to form bright traps, fully opaque regions to form the dark traps, and an intermediate background level to provide an intensity contrast for both type of traps. This mask creates a bright flat-top beam array interleaved with a dark ``flat-bottom" beam array. The beams comprising both arrays are then shaped into Gaussian-like profiles via spatial filtering \cite{Huft2022}. To make the size of the traps suitable for single-atom trapping, the beams are de-magnified and relayed to the atom trapping region using a sequence of lenses in the optical system.  

To enable simultaneous two-species trapping of $^{87}$Rb and $^{133}$Cs at different sites with a single laser, we  select a laser wavelength $\lambda_0$ between the Rb D1 line (795 nm) and the Cs D2 line (852 nm), such that the dynamic polarizabilities of Rb and Cs have opposite signs [Fig. \ref{fig:method_and_design}(b)]. In this scheme, a region of high intensity is attractive to Rb atoms, but repulsive to Cs atoms, and vice versa. We note that there are additional windows of opposing polarizabilities in the $\sim 780 - 790$ nm and $\sim 880 - 895$ nm ranges, however these regions fall near atomic resonances, resulting in high scattering rate and substantially mismatched polarizabilities, potentially causing unequal trap depths between the two species.

To form the appropriate potential gradients, the intensity distribution of flat-top and flat-bottom beams generated with the mask must be smoothed to generate Gaussian-like trapping profiles, which can be accomplished by Fourier-plane spatial filtering with an iris at the intermediate plane of a 4$f$ imaging system \cite{Huft2022}. Smoothing the intensity at the Fourier plane rather than directly on the mask avoids design and fabrication complications, such as the use of grayscale fabrication. Since the dark traps and the bright traps are adjacent to each other, interference between them can result in features that deviate from the desired trapping pattern in the image plane. Fig. \ref{fig:method_and_design}(c) shows the calculated intensity contour of one unit cell in the image plane, for a design that will be described later in this paper. With the selected trapping wavelength $\lambda_0$, Rb atoms are trapped in the bright traps and Cs atoms are trapped in the dark traps. The intensity line profiles of dark and bright traps, shown in Fig. \ref{fig:method_and_design}(d), demonstrate the trapping profiles are both Gaussian-like. Note that we previously used a similar approach to create a large single-species array of Cs atoms in blue-detuned dark 
traps \cite{Huft2022}.

To achieve the best intensity profiles at the image plane for atom trapping, the following requirements should be approached: (a) To maximize the depth of the bright traps, the bright traps should have maximum intensity at the trap sites for a given input power; (b) To maximize the depth of the dark traps, the dark traps should have zero intensity at the trap-site centers; (c) To maximize the energy efficiency of the system, the trapping depth should be equal for the bright and the dark traps for the two atomic species, otherwise more power is required to compensate for the weaker traps.

\begin{figure}[H]
    \centering
    \includegraphics[trim=0pt 0pt 10pt 0pt,clip,width=0.5\linewidth]{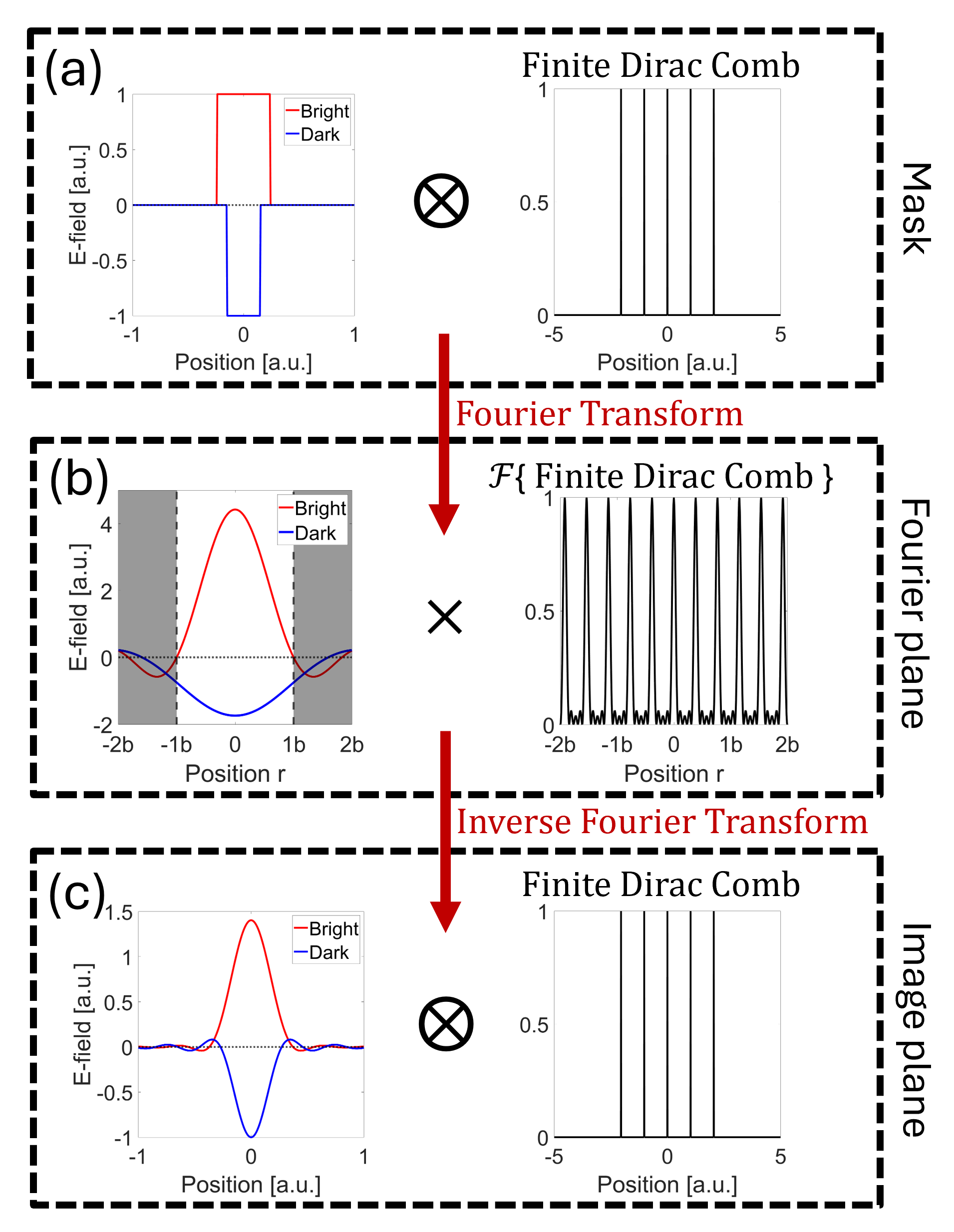} 
    \caption{Gaussian-like trap array generation via Fourier filtering. \textbf{(a) Mask plane:} The field immediately after the array mask in Fig. \ref{fig:method_and_design}(a) can be decomposed into: (left) the electric field of a single transparent or opaque region, convolved with (right) a finite Dirac comb. \textbf{(b) Fourier plane:} A convolution at the mask becomes a product of Fourier transforms. An iris is placed in this plane, blocking the field at $r>b$ and $r<-b$. \textbf{(c) Image plane:} Final filtered electric field of bright and dark traps, obtained via inverse Fourier transform from (b).}
    \label{fig:fourier filtering}
\end{figure}

In our 4$f$ imaging system (Fig. \ref{fig:method_and_design}(a)), the spatial filter is placed at the back focal plane of the first lens (i.e., in the Fourier plane). The amplitude distribution after the optical mask is an array of flat-top/bottom beams, and can be understood as the convolution between the electric field of a single flat-top/bottom (bright/dark) beam with a finite Dirac comb [Fig. \ref{fig:fourier filtering}(a)]. For simplicity, the background amplitude is not discussed here since it is a DC signal which is not filtered by the aperture in the Fourier plane. At the Fourier plane after the first lens (Fig. \ref{fig:fourier filtering}(b)), the electric field can be understood via the Convolution Theorem to be the product of the Fourier transform of the flat-top/bottom beam, and the Fourier transform of the finite Dirac comb. In this plane, we place an iris (gray area in Fig. \ref{fig:fourier filtering}(b)), which is a spatial filter that removes the high-frequency components. At the image plane (Fig. \ref{fig:fourier filtering}(c)), the field profile is the convolution of the inverse Fourier transform of each function in Fig. \ref{fig:fourier filtering}(b)). Note that the spatial filter actually also applies to the Fourier transform of the Dirac comb in Fig. \ref{fig:fourier filtering}(b), but whether or not you apply it is irrelevant because those components end up being multiplied by 0.

The spatial filter is used to remove the high-frequency components to make the trapping profile at the image plane Gaussian-like, which localizes the atoms at the centers of the traps. To achieve the best quality of the traps, certain conditions need to be satisfied, especially the size relationships between the radius of the apertures $a_{\text{bright}}$, the radius of the opaque disks $a_{\text{dark}}$, and the iris radius $b$ [Fig. \ref{fig:method_and_design}(a)].

In the case of bright traps, the beam-amplitude profile after the first lens with focal length $f_1$ becomes a Fourier transform of the flat-top beam, which is $J_{1}(r)/r$, i.e., an airy disk (cross-section shown in Fig. \ref{fig:fourier filtering} (b), $J_{1}(r)$ is the first-order Bessel function of the first kind, and $r$ is the radial distance from the center of each disk). To maximize the depth of the bright traps with a single aperture, the iris radius needs to block the light past the radius where the amplitude goes from positive to zero for the first time, which is shown in Fig. \ref{fig:fourier filtering}(b, left) and Eq. (1): 

\begin{equation}
 b=\frac{f_1}{k a_{\text {bright }}} r_{1}^{(1),}
\end{equation}

\noindent{}where $k$ is the wavenumber of the trapping laser, $f_1$ is the focal length of the first lens, and $r_1^{(1)}$ is the first zero of the first-order Bessel function. This condition ensures that the beam-amplitude profile in the image plane after $f_2$, $A_{\text{bright}}$, is maximized.

Similarly, in the case of dark traps, the amplitude profile at the image plane is the input amplitude $A_0$ minus the complementary Gaussian-like beam profile $A_{\text{dark}}$. The calculation is similar to bright traps but with a key difference: instead of maximizing the center amplitude for bright traps, the center amplitude of dark traps needs to be zero, which means  $A_0-A_{\text{dark}}(r=0)=0$. Since $A_{\text{dark}}(r=0)$ is larger than $A_0$ given the condition in Eq. (1), more light needs to be filtered at the Fourier plane to reduce the amplitude of the complementary flat-top beam down to $A_0$. To achieve this, the size requirement between the dark disk radius $a_{\text{dark}}$ and $b$ is shown in Eq. (2) below, where $r_1^{(0)}$ is the first zero of the zeroth-order Bessel function
\begin{equation}
    b=\frac{f_1}{k a_{\text {dark}}} r_{1}^{(0)}.
\end{equation}
As discussed previously, both bright and dark traps share a single iris in the Fourier plane (Fig. \ref{fig:fourier filtering}(b)). Thus the iris radii in eqs. (1) and (2) are equal, which gives the necessary relationship between the dark and bright trap radii:

\begin{equation}
\begin{aligned}
    \frac{f_1}{k a_{\text{dark}}} r_{1}^{(0)} &= \frac{f_1}{k a_{\text{bright}}} r_{1}^{(1)} \\
    \Rightarrow \quad a_{\text{bright}} &= \frac{r_{1}^{(1)}}{r_{1}^{(0)}} a_{\text{dark}} \approx 1.6 \, a_{\text{dark}}
\end{aligned}
\end{equation}

The arrangement and spacing of the traps depends on the application. As an example, for Rydberg atom-based quantum computing \cite{saffman2010}, atoms are typically spaced at $3 - 10~ \mu \mathrm{m}$ on a square grid, while the traps for single atoms need to have a waist of $\sim 1~  \mu \mathrm{m}$. An appropriate choice is to make  the period about 4 times larger than the size of the apertures for this  application. Other arrangements such as hexagonal clusters\cite{Scholl2021}, Kagome lattices\cite{Semeghini2021}, and triangular lattices\cite{Bedalov2024} have been used for quantum simulation and computation demonstrations. These and other patterns can be readily achieved by using different arrangements of the bright and dark disks on the mask.

To achieve the third requirement -- equal trapping depth for both species atoms -- the intensity depths for dark and bright traps need to satisfy eq. (4) 

\begin{equation}
\begin{aligned}
    \alpha_{\text{dark}} \Delta I_{\text{dark}} &= \alpha_{\text{bright}} \Delta I_{\text{bright}},
\end{aligned}
\end{equation}

\begin{figure}[H]
    \centering
    \includegraphics[trim=5pt 0pt 0pt 0pt,clip,width=1\linewidth]{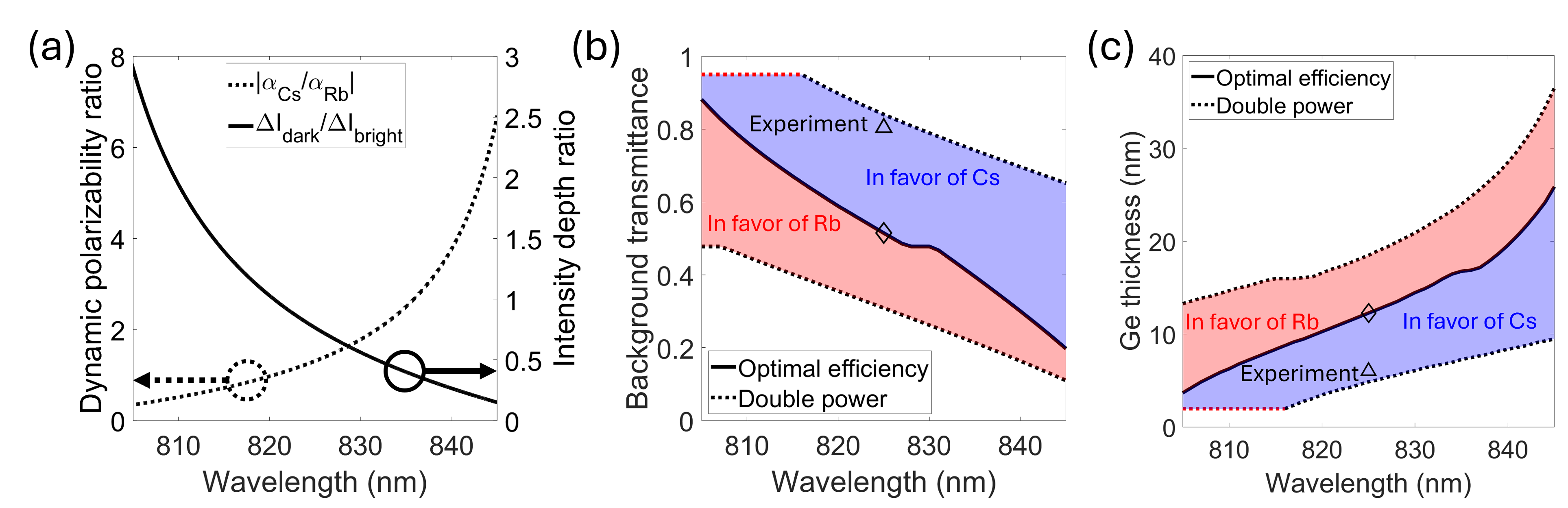}
    \caption{Design of the mask background transmittance. \textbf{(a)} The ground-state dynamic polarizability ratio between Cs and Rb, $|\alpha_{\text{Cs}}/\alpha_{\text{Rb}}|$, decreases as a function of wavelength. To achieve equal trapping depths, we need an intensity-depth ratio,  $\Delta I_{\text{dark}}/\Delta I_{\text{bright}}$, which is the reciprocal of the polarizability ratio. \textbf{(b, c)} The needed (b) background transmittance and (c) the Ge thickness as a function of wavelength, showing the optimal design (solid line), which has equal trapping depth between Cs and Rb, and a range of less-efficient designs (colored regions). The dashed lines indicate designs that require double the incident power compared to the optimum design, to achieve the same trapping depth for the weaker trap. In the blue region, the trapping profile favors Cs atoms (in the dark traps) due to higher-than-optimal background transmittance. Similarly, in the red region, the trapping profile is in favor of Rb atoms (in the bright traps) due to lower-than-optimal background transmittance. Note that when the transmittance is very high (i.e., Ge thickness is very thin), the interference ripples (shown in Fig. 1(c)) start to be comparable to the bright traps. Due to this, we set a cap of the transmittance at 95\% and a bottom of the Ge thickness at 2 nm to avoid the bright-trap intensity depth being smaller than the interference ripples. The triangles mark the experimentally demonstrated design. The diamonds mark the optimal design for a trap laser of wavelength 825 nm.}
    \label{fig:wavelength_vs_transmittance}
\end{figure}

\noindent{}where $\alpha$  is the ground-state polarizability of the atoms (shown in Fig. 1(b)) and $\Delta I$ is the intensity depth of the traps (shown in Fig. 1(c)). This requirement helps the mask to create effective dark and bright traps with the lowest input laser power. To control the intensity depths, we use a semi-transparent background layer to provide contrast for both bright and dark traps. If a longer wavelength is used, the required intensity depth ratio between dark and bright traps decreases, decreasing the required background transmittance, as shown in Fig. \ref{fig:wavelength_vs_transmittance}(b). 

Notably, our approach is robust to errors in the background transmittance. When the background transmittance is not optimal, both species of atoms can still be trapped, but at a cost of higher laser power to compensate for the weaker traps. For example, if we are willing to double the minimum required power, a range of background transmittance can be accepted, which is shown in Fig. \ref{fig:wavelength_vs_transmittance}(b). Higher background transmittance favors the dark traps (here, Cs atoms), while lower background transmittance favors the bright traps (here, Rb atoms). 

\section{Materials and non-ideal phases}

\begin{figure}[H]
    \centering
    \includegraphics[trim=16pt 0pt 0pt 0pt,clip,width=0.63\linewidth]{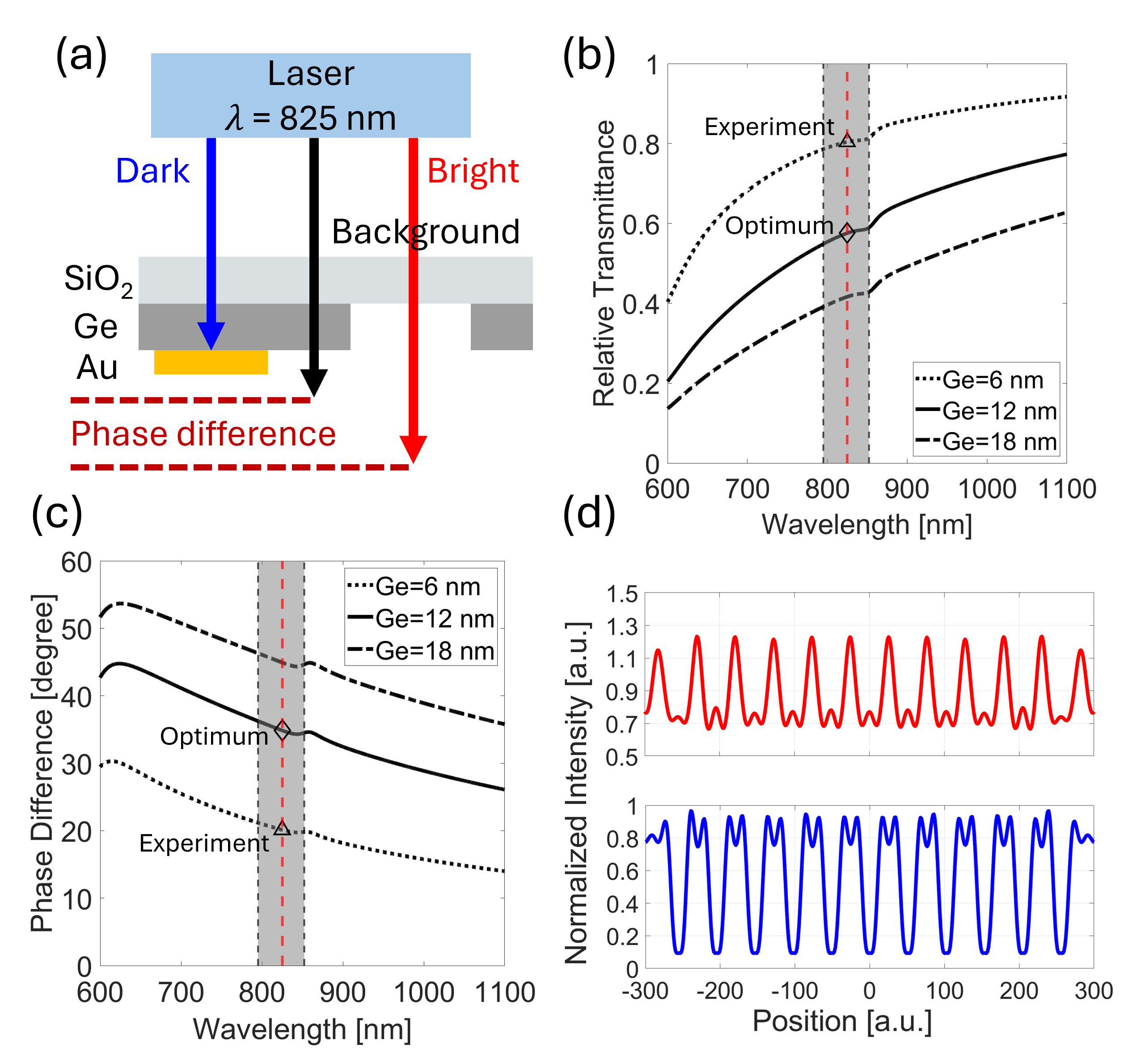}
    \caption{Design of the mask transmittance and phase. \textbf{(a)} The mask has three levels of transmittance: (1) The dark traps are produced by circular regions coated with optically thick (opaque) gold; (2) the bright traps are produced by circular open apertures, which are just the bare AR-coated substrate; and (3) the background regions has an intermediate transmittance level, here generated by a 6-nm-thick layer of evaporated amorphous germanium (a-Ge). The high absorption coefficient of a-Ge ensures a small phase difference between the background and the bright region. \textbf{(b)} The simulated relative transmittance between the background and the bright regions for a-Ge thicknesses of 6 nm, 12 nm, and 18 nm of Ge.  \textbf{(c)} The simulated phase difference between the background and the bright regions. In both (b) and (c), at wavelength = 825 nm, the triangle marks the phase difference of the experimentally demonstrated design and the diamond marks the optimal design. \textbf{(d)} The simulated intensity profiles of both dark (blue) and bright (red) traps at $\lambda_0 = $ 825 nm when taking the phase difference into account for the optimum case. Compared to the optimal line profile in Fig. \ref{fig:method_and_design}(d), both the dark and bright traps are slightly shallower.}
    \label{fig:phase and transmittance}
\end{figure}

On the intensity mask, there are regions with three different levels of transmittance $T$ (shown in Fig. \ref{fig:phase and transmittance}(a)): (1) transparent regions ($T_{\text{bright}} = 1$) for generating the bright traps; (2) a background ($0 < T_{\text{background}} < 1$); and (3) an opaque region ($T_{\text{background}} = 0$) for generating dark traps. Additionally, an important requirement is that there is no substantial phase difference between the background and the bright regions, which creates additional constraints on how the background region should be designed.

To make the highly transparent regions, we used a window substrate with anti-reflection (AR) coatings on both sides. For the opaque regions, we deposited 100 nm of gold, which results in a transmittance less than 0.1\%. 

To achieve the desired transmittance in the background region, we used a thin layer of material with an intermediate degree of absorption such that (a) a subwavelength-thick layer can result in enough absorption of the incident light without too much phase shift, and (b) the absorption is not so high, such that the transmittance can be controlled on the single-percent level by changing deposition conditions such as deposition time. We found that amorphous semiconductors with band gaps smaller than the photon energy, such as amorphous germanium (a-Ge), can work in this role. Here, we chose electron-beam evaporated Ge, which can be smooth for thin layers \cite{Ciesielski2018}. We used the transfer-matrix method (TMM) to calculate the transmittance and phase shift in the background region coated with a-Ge, and the results are shown in Fig. \ref{fig:phase and transmittance}(b, c). The wavelength-dependent refractive index and extinction coefficient are adapted from Ref.~\cite{Palik1998}.

Based on the calculated relative transmittance and phase difference between the background the transparent region, we performed a simulation similar to Fig. \ref{fig:fourier filtering}, consisting of a Fourier transform of the electric field after the mask to yield the field distribution in the Fourier plane, spatial filtering at the Fourier plane, then an inverse Fourier transform to find the field distribution at the image plane. In the previous simulation, zero phase difference between the background and the transparent region was assumed. Now we include the accumulated phase in the transparent region, $\phi_{Air}$, and in the background region, $\phi_{Ge}$, and the resulting complex-field spatial distribution is then fed into the calculation described in Fig. \ref{fig:fourier filtering}. This simulation considers interference between the dark and bright traps, which enables us to precisely calculate what a-Ge thickness is required. When the background material thickness is not optimal, this mask can still be used to trap both species of atoms, but requires using a higher laser power. 

In our design, using a 825 nm laser, $^{87}$Rb atoms are trapped in the bright traps and $^{133}$Cs atoms are trapped in the dark traps. The ground-state dynamic polarizability ratio of Cs and Rb is about 2 for the design wavelength $\lambda_0=825 \textrm{ nm}$, so the intensity depth ratio of dark and bright traps is around 0.5 (in Fig. \ref{fig:wavelength_vs_transmittance}(a)). Assuming that we can accept doubling the optimal laser power, a Ge thickness from 5 nm to 18 nm can trap both species of atoms. 

We evaporated an a-Ge layer, with thickness of approximately 6 nm, to form the background region, which according to our TMM calculation has a transmittance of ~0.8 for $\lambda = 825$ nm and a phase difference of 20$^\circ$ compared to the transparent region [Fig. \ref{fig:phase and transmittance}(b, c)]. This phase difference is small but not zero, and introduces a more complicated interference pattern in the image plane compared to the case of no phase difference, which leads to shallower trapping depths for both dark and bright traps. A simulated cross-sectional intensity profile accounting for the phase difference is shown in Fig. \ref{fig:phase and transmittance}(d). 

The sizes of the dark and bright traps at the back focal plane of lens 2 can be controlled by the focal length ratio between the first lens $f_1$ and the second lens $f_2$. Thus, the bright and dark aperture sizes on the intensity mask can be many times larger than the trap sizes, which are usually micron-scale. We chose the radii of the bright disks to be 120 $\mu \mathrm{m}$ and dark disks to be 75 $\mu \mathrm{m}$ (Table \ref{tab:mask-params}), which is trivially achieved with contact photolithography or direct laser writing.

\begin{table}[H]
    \centering
    \begin{tabular}{|>{\centering\arraybackslash}m{0.2\linewidth}|>{\centering\arraybackslash}m{0.2\linewidth}|>{\centering\arraybackslash}m{0.2\linewidth}|>{\centering\arraybackslash}m{2cm}|}
\hline
      $a_{\rm bright}$  &  $a_{\rm dark}$  & $\rm Period$  &  Ge thickness\\
\hline
      $120~ \mu \mathrm{m}$  &  $75~ \mu \mathrm{m}$  &  $516~ \mu \mathrm{m}$  &  $6~ \mathrm{nm}$\\
\hline
\end{tabular}
    \caption{Physical parameters of fabricated mask.}
    \label{tab:mask-params}
\end{table}

\section{Fabrication and characterization}

\label{sec:III fabricaiton and characterization}
\begin{figure}[H]
    \centering
    \includegraphics[trim=0pt 0pt 0pt 13pt,clip,width=0.8\linewidth]{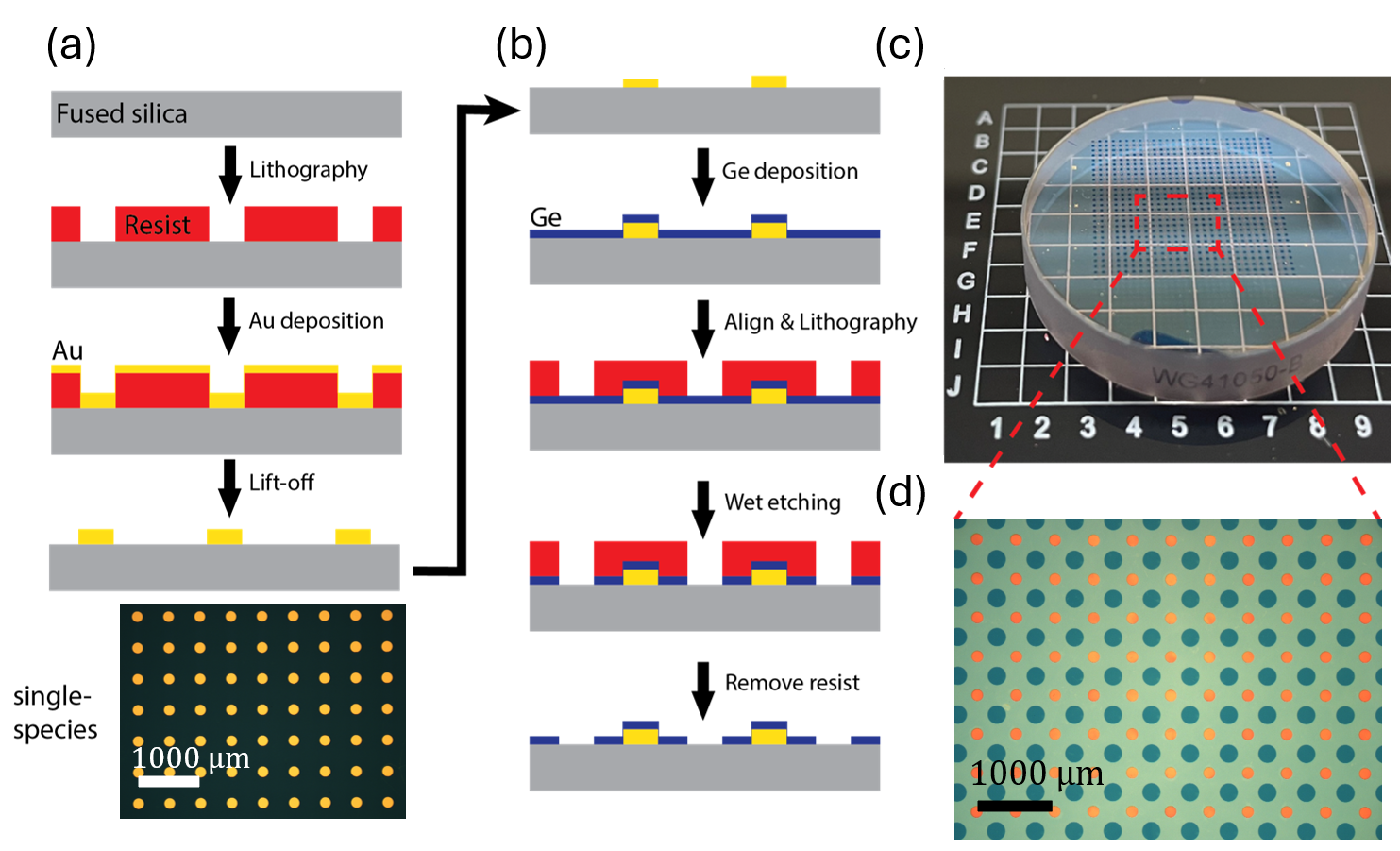}
    \caption{The fabrication processes of the two-species optical mask. \textbf{(a)} The dark regions made of 100 nm gold thin films are fabricated by laser lithography, electron-beam evaporation, and lift-off. \textbf{(b)} The background regions, comprising 6 nm Ge, are formed by Ge evaporation, lithography with alignment, and etching of Ge. \textbf{(c)} The two-species optical mask on a one-inch glass substrate with double-side anti-reflection coating. \textbf{(d)} Microscope image of the mask. The distance between adjacent dark disks is 512 $\mu \mathrm{m}$, the radius of the bright disks is 120 $\mu \mathrm{m}$ and the radius of the dark disks is 75 $ \mu \mathrm{m}$.}
    \label{fig:fabrication process}
\end{figure}

The fabrication process can be separated into two broad steps: the first, presented in Fig. \ref{fig:fabrication process}(a), results in a mask that can be used for single-species trapping, and then the second step, presented in Fig. \ref{fig:fabrication process}(b), results in the mask for a two-species trap array.

First, we fabricate the gold disks for dark trapping using a lift-off photolithography process with S1813 resist. After resist development, we evaporate 2 nm of Ti and 100 nm of Au, and perform liftoff. Then, we evaporate the amorphous Ge thin film, resulting in the background region (Fig. \ref{fig:fabrication process}(b)). Note that the Ge is also deposited on top of the gold regions, but this does not change the transmittance, which is already zero. Then we spin another layer of photoresist, and use laser lithography with alignment to expose the regions that will become the bright traps. The pattern of bright traps is aligned to be in between dark traps with spacing determined by the needs of an atomic experiment or device. After exposure and development, the Ge is etched away using a hydrogen peroxide wet etching process \cite{Huygens2007}, followed by removal of the remaining resist.

Fig. \ref{fig:fabrication process}(c, d) is a photo and a microscope image of the fabricated two-species mask on a one-inch AR-coated window. The transmittance of the background region was measured over 4 months in standard lab atmosphere to make sure that it does not change due to oxidation of Ge \cite{Ligenza1960}. 

\section{Trapping Rb and Cs atoms}
\label{sec: III trapping atoms}

\begin{figure}[H]
    \centering
    \includegraphics[trim=5pt 0pt 0pt 5pt,clip,width=0.5\linewidth]{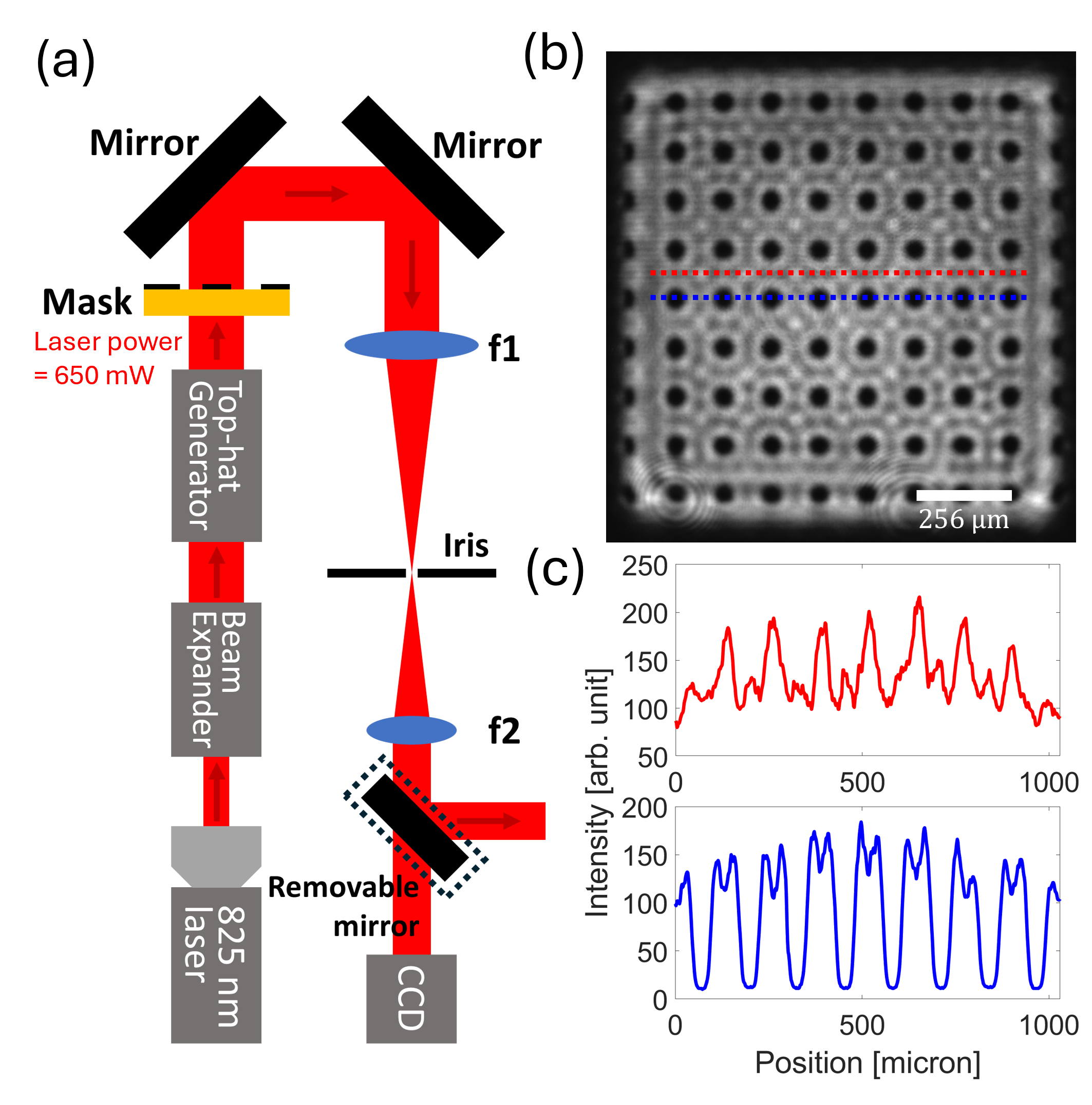}
    \caption{Optical setup. \textbf{(a)} The 825 nm laser is beam-expanded to fill the region on the optical mask that generates the traps shown in (b). To achieve more uniform traps, a diffractive top-hat beam generator is used to produce a square uniform beam profile. The laser power before the mask is approximately 650 mW. The output of the mask is imaged by a CCD camera after $4\times$ de-magnification, while also passing through the spatial filter (iris). A removable mirror is used to image a plane conjugate to the trap in the vacuum cell, as shown in Fig. 5. \textbf{(b, c)} The measured (b) two-dimensional and (c) line profiles of the intensity at the CCD camera.}
    \label{fig:optical measurement}
\end{figure}

In our experiment, the mask was illuminated with an 825 nm diode laser amplified by a tapered amplifier (MOGLabs). The incident beam was diffracted using an acousto-optic modulator for intensity control and switching, and then coupled into a single-mode optical fiber. After the intensity mask, the beam was de-magnified by a 4$f$ optical system with $f_1=250$~mm and $f_2=125$ mm and shaped by a variable iris on an x-y translation stage mounted in the Fourier plane. The optical setup is shown in Fig. \ref{fig:optical measurement}(a), where a top-hat generator was also used to form a uniform intensity profile. A CCD camera is placed at the focal plane of lens 2 to monitor the shape of the trapping beams before further de-magnification into the vacuum cell. The iris was centered on the beam by closing it all the way and optimizing the x-y symmetry of the resulting pattern. Then the iris was opened and adjusted while monitoring the image to maximize the brightness of the bright traps. Fig. \ref{fig:optical measurement}(b) shows an image of the trap array on the CCD camera. Fig. \ref{fig:optical measurement}(c) shows the Gaussian-like line profiles of the dark and bright traps highlighted in Fig. \ref{fig:optical measurement}(b) near the center of the array. Due to higer-than-ideal transmittance in the background region, the trap depths of the dark traps are deeper than the bright traps, but this can be addressed by adjusting the background transmittance level.

\begin{figure}[H]
    \centering
    \includegraphics[trim=10pt 0pt 0pt 5pt,clip,width=1\linewidth]{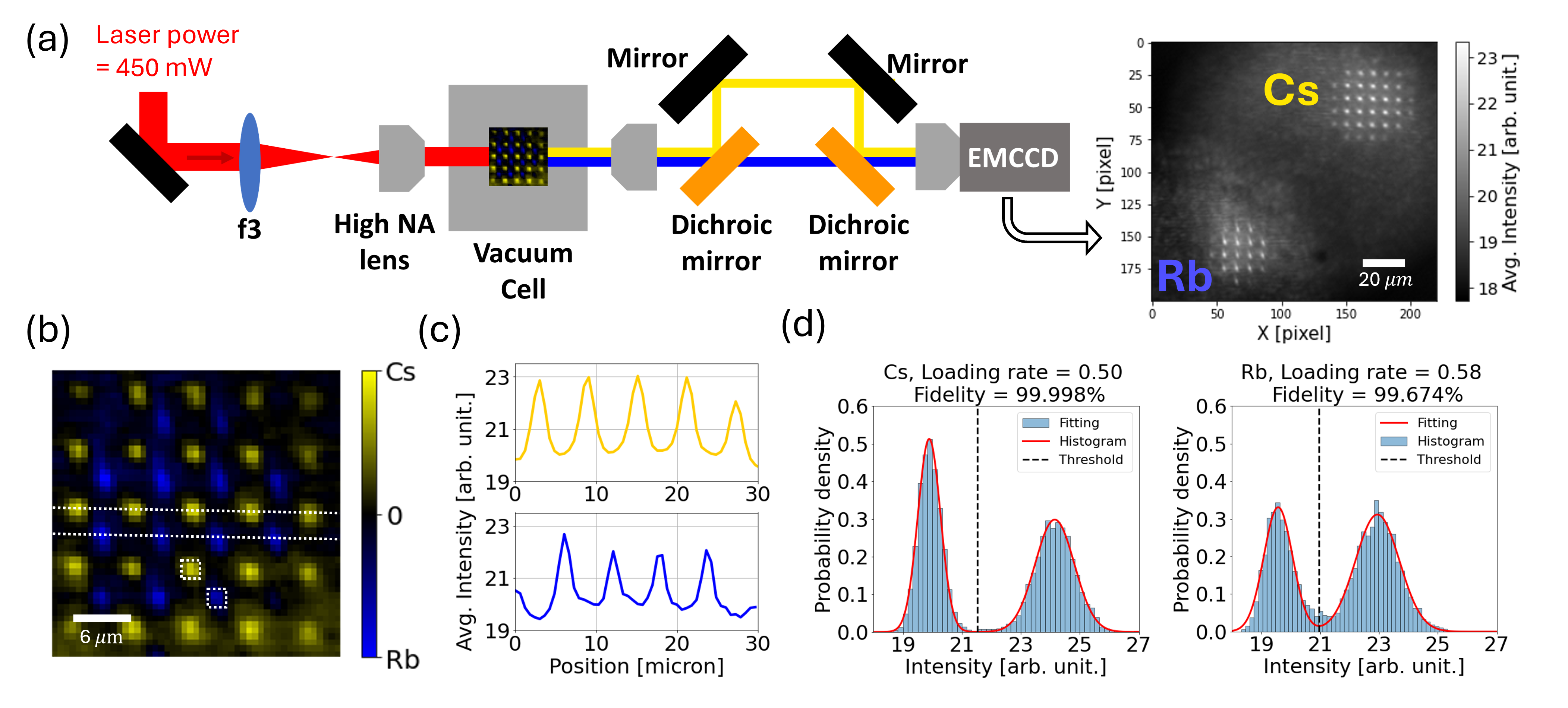} 
    \caption{Two-species atom trapping. \textbf{(a)} The laser power after the mask and the Fourier filtering is 450 mW. After $f_3$ and the high-NA lens, the plane in Fig. \ref{fig:optical measurement}(b) is re-imaged into the vacuum cell with further de-magnification, to form the atom traps. On the other side of the vacuum cell, the fluorescence from Rb and Cs atoms is spatially separated to form two fluorescence images on a single EMCCD. The Rb array is imaged at the lower left, and the Cs array is at the upper right of the EMCCD. \textbf{(b)} The reconstructed averaged images of the Cs and Rb single-atom arrays, with fluorescence from Cs in yellow and Rb in blue. \textbf{(c)} The intensity profiles of the Rb and Cs fluorescence at the specific lines highlighted in (b). \textbf{(d)} Histograms of the  EMCCD photoelectron counts from individual Cs and Rb sites over 5000 loading shots. The individual Cs and Rb sites are highlighted in (c). The two peaks correspond to the measured intensity from sites with zero atoms, and sites with one atom.}
    \label{fig:atom trapping expriment}
\end{figure}

After the intermediate-step imaging, a mirror was inserted in the beam path, then a sequence of lenses with focal lengths $f_{3-6}=(200, -50, 250, 23.2) \text{ mm}$ was used to relay the transmitted light into the vacuum cell for atom trapping (shown in Fig. \ref{fig:atom trapping expriment}(a)). However, the top-hat beam generator is removed before further experiment for simplicity. The $23.2$~mm lens is a custom objective lens with high numerical aperture (NA) and correction for the vacuum cell wall, suitable for creating a diffraction-limited image of the trap array at the center of a glass vacuum cell where $^{87}$Rb and $^{133}$Cs were collected in a dual-species 3D magneto-optical trap (MOT). The de-magnification from the mask to the atoms was $86.2\times$, corresponding to a periodicity of $5.94~\mu$m with a total power of 450 mW. Considering the input laser power is approximately 650 mW, the power efficiency of this specific imaging system is about $69\%$. 

Atoms were first loaded into a dual-species MOT. The magnetic quadrupole gradient was abruptly turned off and the detuning of the cooling light from resonance was increased to further cool the atoms via polarization gradient cooling (PGC)  \cite{Dalibard1989}. During the PGC phase, atoms were stochastically loaded into the optical tweezer (bright) and bottle (dark) traps, where they underwent collisional blockade \cite{Schlosser2002} until at most one atom per trap site survived. The resulting atom arrays were then imaged by pulsing quasi-resonant light from the side of the vacuum cell (shown in Fig. \ref{fig:atom trapping expriment}(a)) to induce fluorescence at wavelengths 780 nm for Rb and 852 nm for Cs. This fluorescence was collected via a second high-NA objective on the other side of the vacuum cell. A series of dichroic beamsplitters was used to separate the trapping light from the fluorescence and to split the fluorescence light into two separate imaging paths, one for each species, before recombining the images of the two interleaved arrays onto a single electron-multiplying charge-coupled device (EMCCD, Nuvu HN\"{u}512). The imaging separation of Rb and Cs atoms helps increase the signal to background of each species. The averaged fluorescence image of 5000 loading shots collected by the EMCCD is shown in Fig. \ref{fig:atom trapping expriment} (a) with the Rb array of atoms in the lower-left and the Cs array in the upper-right. The exposure time of each fluorescence measurement is 50 ms. Each atom-trapping site is spatially separated by about 6 $\mu \mathrm{m}$. The atom lifetime in the traps was observed to be seconds to tens of seconds.

In the averaged fluorescence image, both Cs and Rb arrays have background some fluorescence around them. We understand this background to consist of three components: (1) A large portion is due to scattering the 852 nm and 780 nm probe light off of the cell that makes it to the camera. We observed that when the MOT is off, the EMCCD still collects substantial light from the probe lasers. Also, when we block part of the probe laser, shadows show up on the EMCCD; (2) additional background is due the atoms trapped at the Talbot planes (Fig. \ref{fig:talbot effect}). When we scanned the objective lens in the z direction, we observed atoms trapped in planes other than the image plane; (3) for the Rb array, some background light is caused by the fluorescence of the atoms being trapped in the Gaussian envelope (Fig. \ref{fig:intensity profile with gaussian input}(a, b)) created by the input Gaussian beam. The Gaussian envelope behaves as a much larger bright trap. This background noise can be mitigated by using a top-hat beam instead of Gaussian beam. As a result, the fluorescence-based readout of Rb atoms generally has lower readout fidelity in our trapping experiment, which we will discuss below.

In Fig. \ref{fig:atom trapping expriment}(b), we reconstructed the atomic fluorescence images of Cs and Rb by aligning the centers of each array. As per the design, the two arrays of different species are interleaved. This image is displayed by defining the intensity values of Cs as positive and the intensity values of Rb as negative, then summing them. Cs atoms are shown as yellow [RGB = (1, 1, 0)], and Rb atoms are shown as blue [RGB = (0, 0, 1)]. The line profiles of the fluorescence from Cs and Rb atoms are shown in Fig. \ref{fig:atom trapping expriment}(c). 
In Fig. \ref{fig:atom trapping expriment}(d), the atomic fluorescence histograms of Cs and Rb at selected sites are plotted. The fluorescence histogram is fitted with a sum of two Gaussian distributions, one centered at the background signal intensity level and the other centered at a higher intensity level, indicating a single atom. Once the double-Gaussian model has been fitted, the intensity at the lowest probability density level between the two peaks is set as the threshold intensity (dashed lines in Fig. \ref{fig:atom trapping expriment}(d)). When the intensity is higher than the threshold, the trap is considered to be occupied, and vice versa. Based on the fitting parameters and the threshold intensity, we calculated the readout fidelity \cite{Shea2020}, which is listed on top of Fig. \ref{fig:atom trapping expriment}(d). For the entire array of both Rb and Cs atoms, the fitting of the histograms and the calculated readout fidelities for each trapping site are plotted in supplementary Fig. \ref{fig:Cs array histogram} and Fig. \ref{fig:Rb array histogram}.

\section{Conclusion}

 We demonstrated simultaneous trapping of individual Rb and Cs atoms, positioned in an atom-specific interleaved grid, using a single trapping laser and an optical system comprising a three-transmittance-level optical mask and a spatial filter. This is a simpler approach to trapping arrays of multiple atomic species compared to the use of active optical components like SLMs, AODs, and DMDs, and multiple lasers. The array generator works for a relatively wide range of wavelengths, is robust to fabrication imperfections, and can be scaled to arbitrarily large array sizes by scaling the laser power with the number of desired traps. The approach can be readily adapted to any desired configuration of atom traps at any wavelength, and will therefore be useful for a variety of single-atom experiments and applications. 
 
 Even though atom loading is stochastic across the array, the simultaneous use of dark and bright traps has the useful feature that Rb atoms are only collected in bright traps and Cs atoms are only collected in dark traps. This reduces the complexity of atom rearrangement for preparing fully occupied, interleaved arrays. Although the use of fabricated components reduces operational complexity compared to active pattern forming devices, such components do not easily allow for correction of optical imperfections that may lead to trap depth variation across the array. Minimization of such variations is important for applications such as optical tweezer clocks\cite{Madjarov2019,Young2020clock}. Development of hybrid systems that rely on passive components for the major part of beam shaping while incorporating the ability to make smaller adjustments for improved array uniformity is an interesting direction for future work.   

\section*{Competing Interests}

CF, SD, PH, MAK, and MS are inventors on a patent application (P240121US01) submitted by the Wisconsin Alumni Research Foundation (WARF), covering devices and methods to trap arrays of isolated particles of multiple species. All other authors declare they have no competing interests.
 
\section*{Funding}

This material is based upon work supported by the U.S. Department of Energy Office of Science National Quantum Information Science Research Centers as well as support from NSF Award 2016136 for the QLCI center Hybrid Quantum Architectures and Networks, and NSF award 2210437.

\section*{Author Contributions}

Conceptualization: CF, MAK, PH, MS. Calculations and simulations: CF, SD, SAN, PH. Experimental realization: CF, JM, JG, ML, MG, MB, SAN. Data analysis: CF, JM, JG, MAK, MS
Project administration: MAK, MS. Writing original draft: CF, SAN, MAK, MS. Review and editing: all authors.

\section*{Data and Materials Availability}

Data underlying the results is openly available in a Zenodo repository at \href{https://doi.org/10.5281/zenodo.14533096}{doi.org/10.5281/zenodo.14533096}.

\bibliographystyle{unsrt}   
\bibliography{references,qc_refs,rydberg,saffman_refs,atomic}

\newpage
\section*{Supplementary Materials}
\subsection*{Ge thin film transmittance stability}

To evaluate the long-term stability of the evaporated Ge thin film's optical properties, especially the transmittance, we measured the relative transmittance of the Ge-coated background region versus the transparent region after four months of using the optical mask. The optical mask is used in an atmospheric environment with 808 nm laser exposure at around 1 W. As shown in Fig. \ref{fig:transmittance after four month}, no substantial changes in transmittance are observed, indicating that the Ge thin film maintains its optical properties under high-power laser exposure.

\setcounter{figure}{0}
\renewcommand{\figurename}{Figure}
\renewcommand{\thefigure}{S\arabic{figure}}

\begin{figure}[H]
    \centering
    \includegraphics[width=0.4\linewidth]{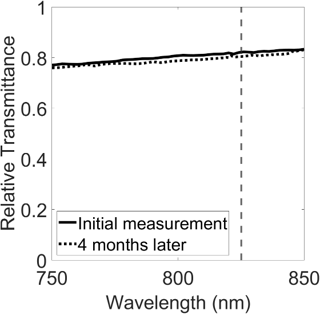} 
    \caption{The relative transmittance between the background region (with Ge thin film) and the transparent region at wavelengths from 750 nm to 850 nm.}
    \label{fig:transmittance after four month}
\end{figure}

\newpage
\subsection*{Escape position of the traps }
\begin{figure}[H]
    \centering
    \includegraphics[trim=0pt 0pt 0pt 10pt,clip,width=\linewidth]{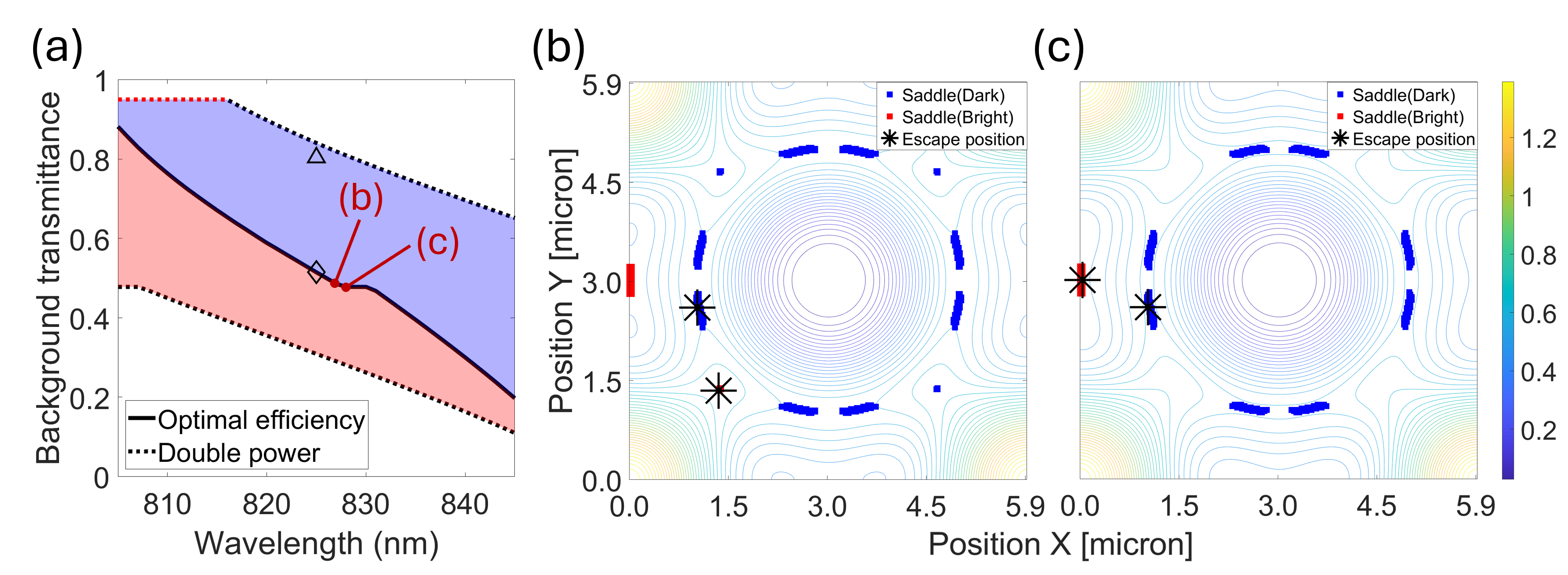} 
    \caption{Escape point of the bright and dark traps. \textbf{(a)} Reproduced figure from Fig. \ref{fig:wavelength_vs_transmittance}, highlighting two particular wavelengths: 826 nm and 827 nm. \textbf{(b)} The potential energy contour at a wavelength of 826 nm. \textbf{(c)} The potential energy contour at a wavelength of 827 nm.}
    \label{fig:Escape position}
\end{figure}

\newpage
\subsection*{Cs array histogram}

\begin{figure}[H]
    \centering
    \includegraphics[width=\linewidth]{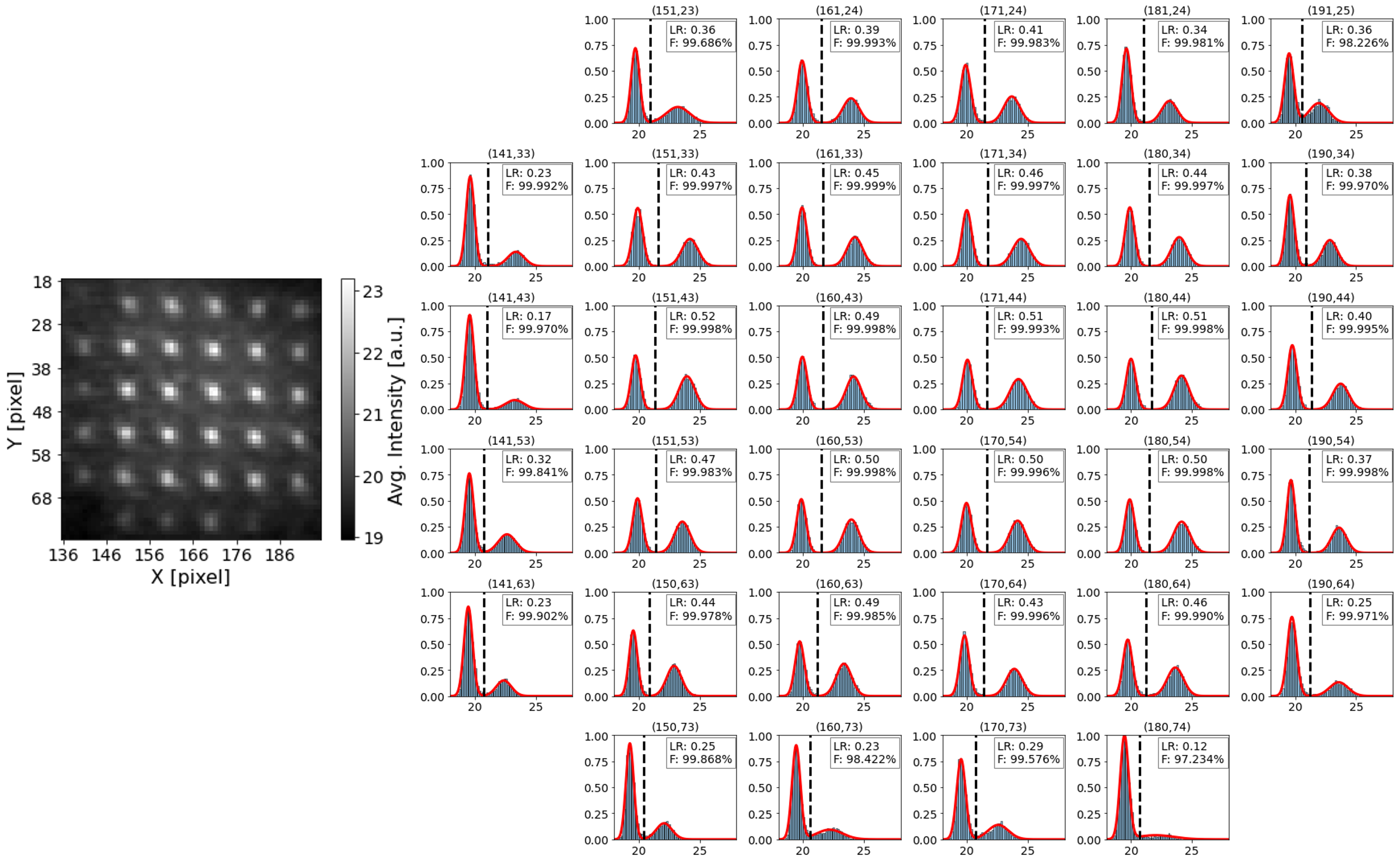} 
    \caption{Fluorescence readout of Cs atom array. (left) An averaged EMCCD image of 5000 individual shots. (right) The histograms of the EMCCD photoelectron counts from individual Cs sites. A double-Gaussian model (red curve) is used to fit the histogram. LR is the loading rate of a single shot and F is the readout fidelity.}
    \label{fig:Cs array histogram}
\end{figure}

The fluorescence-based readout of Cs atoms at each trapping site across the entire array is shown in Fig.\ref{fig:Cs array histogram}. Double-Gaussian distribution model is used to fit the histograms. The fitting parameters are used to determine the threshold intensity (vertical dashed line in Fig.\ref{fig:Cs array histogram}), then calculate the loading rate and the readout fidelity.

\newpage
\subsection*{Rb array histogram}

\begin{figure}[H]
    \centering
    \includegraphics[width=\linewidth]{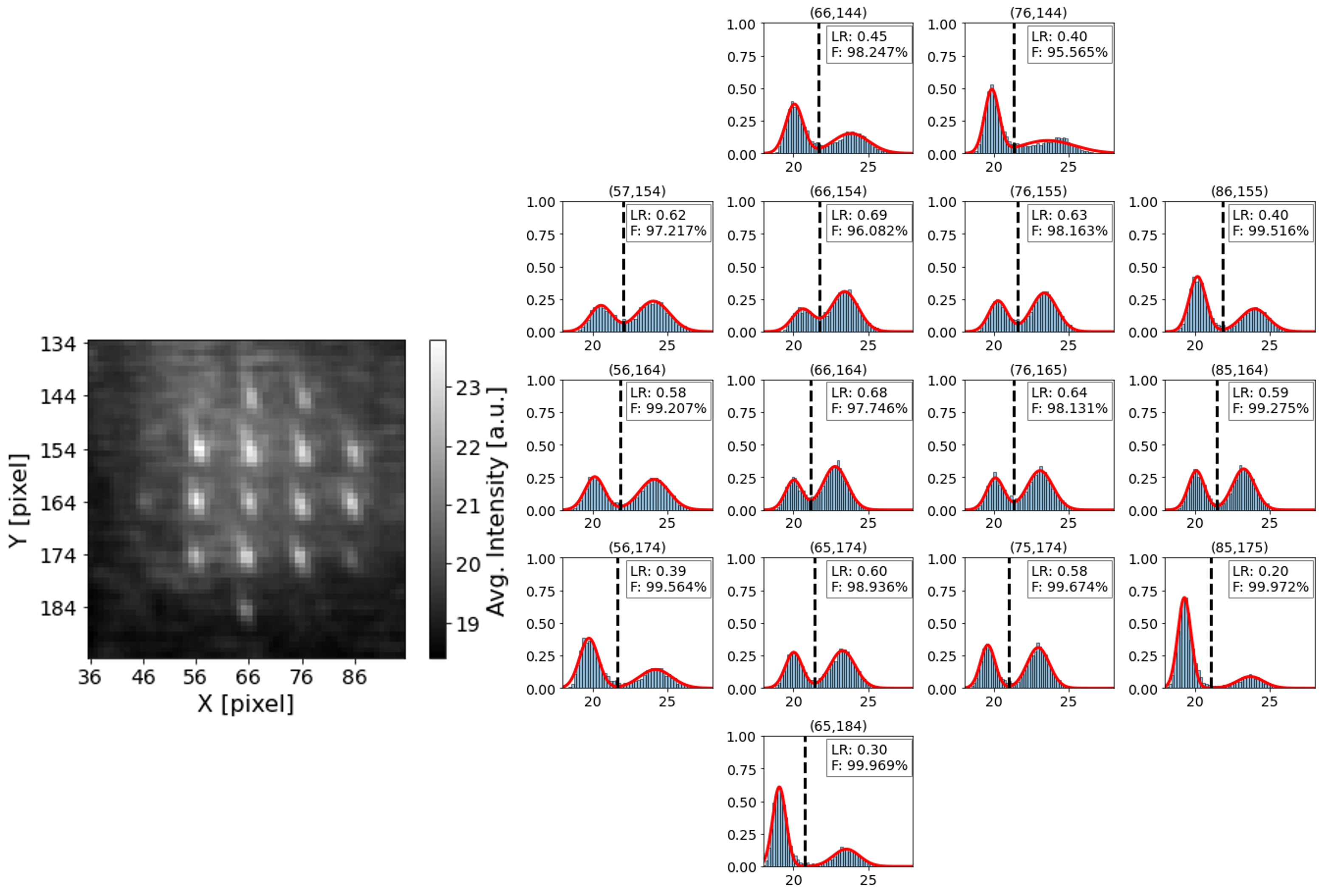} 
    \caption{Fluorescence readout of Rb atom array. (left) An averaged EMCCD image of 5000 individual shots. (right) The histograms of the EMCCD photoelectron counts from individual Rb sites. A double-Gaussian model (red curve) is used to fit the histogram. LR is the loading rate of a single shot and F is the readout fidelity.}
    \label{fig:Rb array histogram}
\end{figure}

The fluorescence-based readout of Rb atoms at each trapping site across the entire array is shown in Fig.\ref{fig:Rb array histogram}. The fitting parameters are used to determine the threshold intensity (vertical dashed line in Fig.\ref{fig:Rb array histogram}), then calculate the loading rate and the readout fidelity.

\newpage
\subsection*{Imaging the dual-species array with less coma aberration}

\begin{figure}[H]
    \centering
    \includegraphics[width=.5\linewidth]{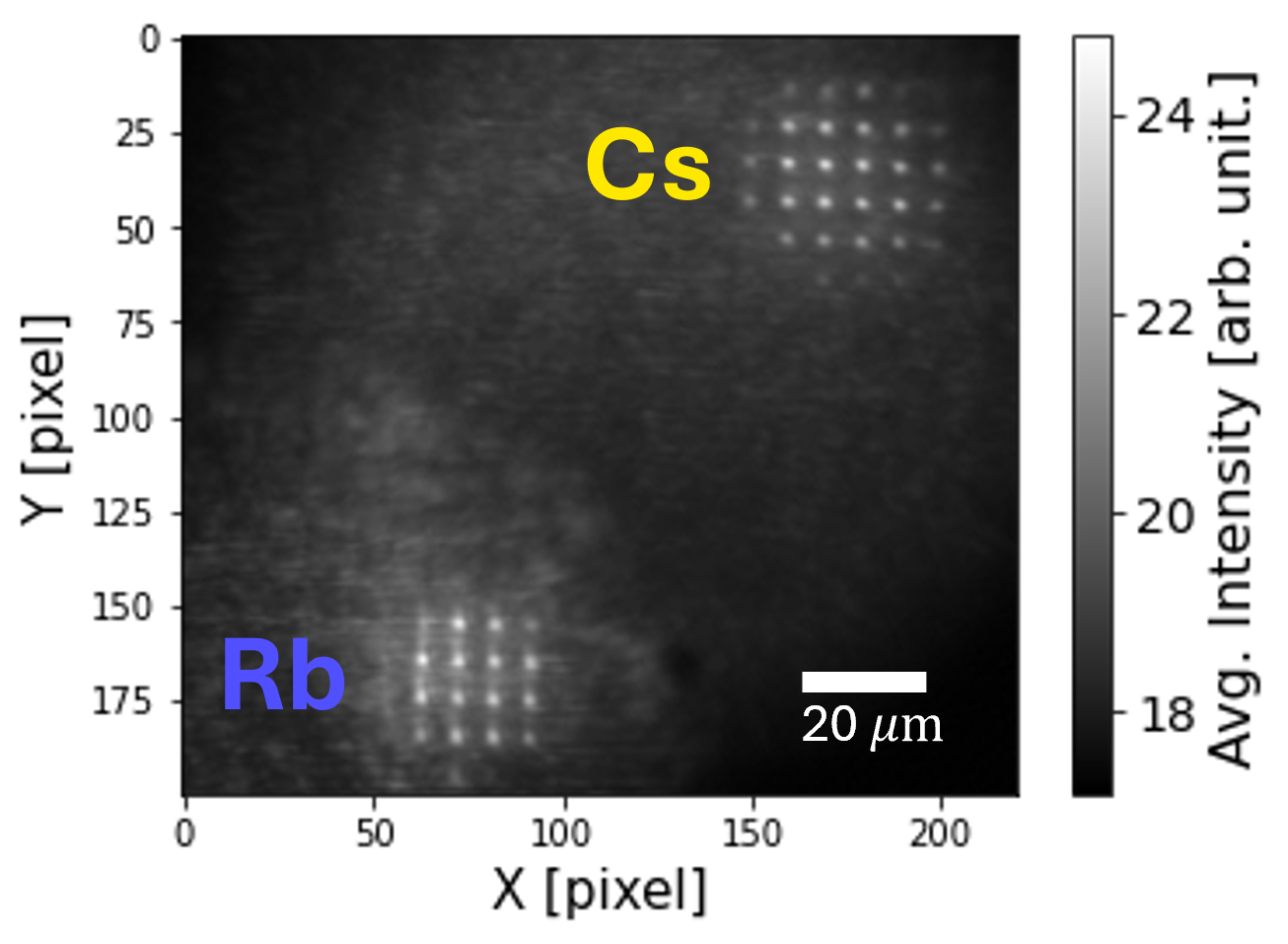} 
    \caption{The averaged fluorescence image of 2000 loading shots. The Rb array and the Cs array are spatially separated. The Rb array is at the lower left, and the Cs array is at the upper right of the EMCCD.}
    \label{fig:better looking array}
\end{figure}

In Fig. \ref{fig:atom trapping expriment}(a), the averaged fluorescence image of the Rb array is distorted. This is because the imaging path of Rb array is slightly off axis. In the fluorescence readout process, we spatially separate the Cs and Rb arrays using dichroic beam splitters. However, this introduces coma aberration on the Rb array. To verify that the distortion can be fixed by better optical alignment or a reduction of the aberrations, another set of data was collected (Fig. \ref{fig:better looking array} with 2000 loading shots. The Rb array shows less coma aberration.

\newpage
\subsection*{Talbot effect}

\begin{figure}[H]
    \centering
    \includegraphics[width=.5\linewidth]{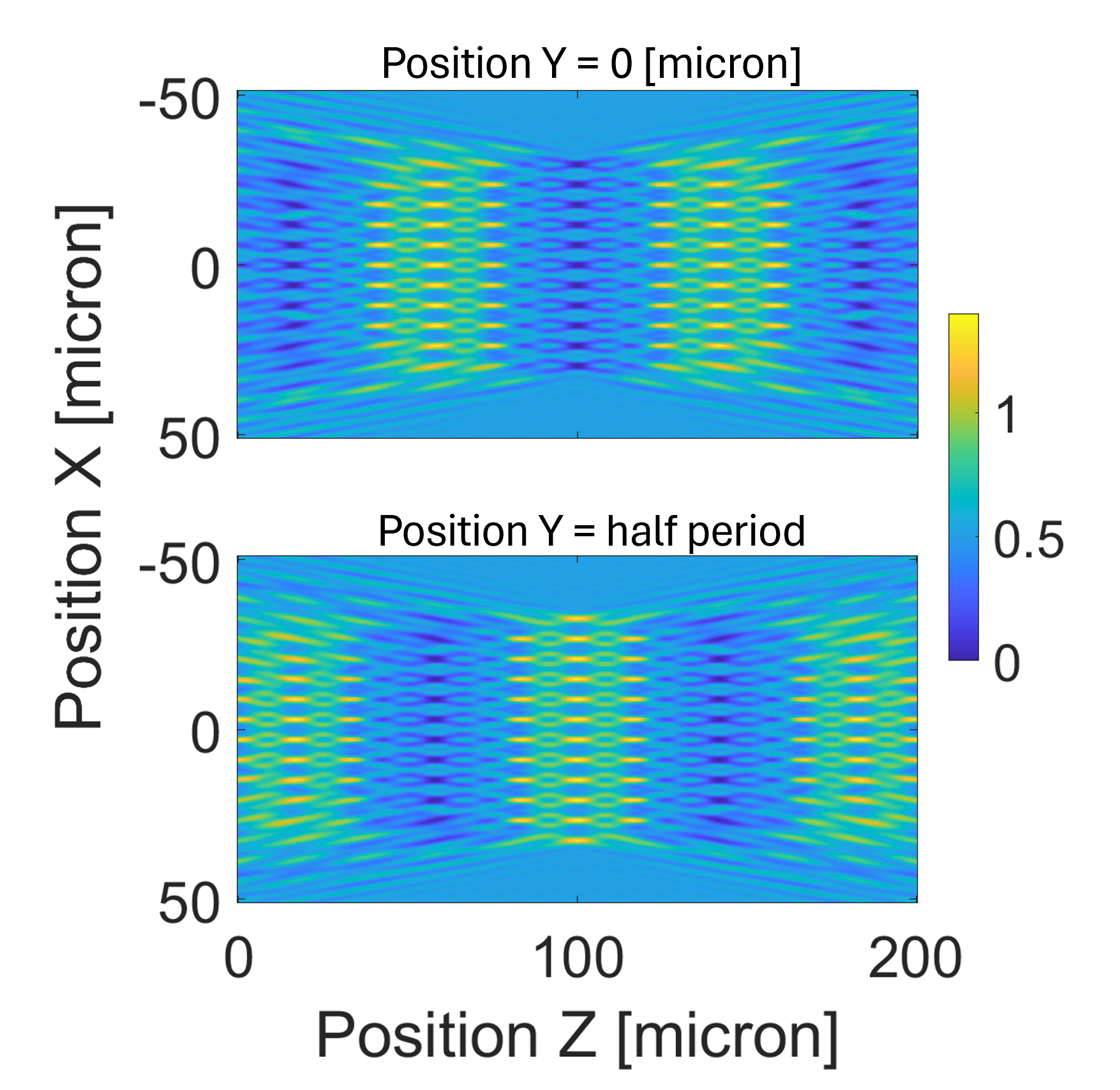} 
    \caption{The Talbot plane near the image plane of the trapping system. The image plane is located at the middle of the Z direction. The top figure shows the Talbot planes for dark traps, while the bottom figure shows Talbot planes for bright traps.}
    \label{fig:talbot effect}
\end{figure}

Since the trapping profile has periodic structure, the Talbot effect causes self-images of the intensity pattern at the light propagation direction, shown in Fig. \ref{fig:talbot effect}. This unintended self-imaging might blur fluorescence images. In our trapping experiment, we have noticed the atom loading at Talbot planes when the objective lens is scanned in Z direction.  

\newpage
\subsection*{Intensity profile with Gaussian input and resulting light shift}

\begin{figure}[H]
    \centering
    \includegraphics[trim=15pt 0pt 0pt 15pt,clip,width=.8\linewidth]{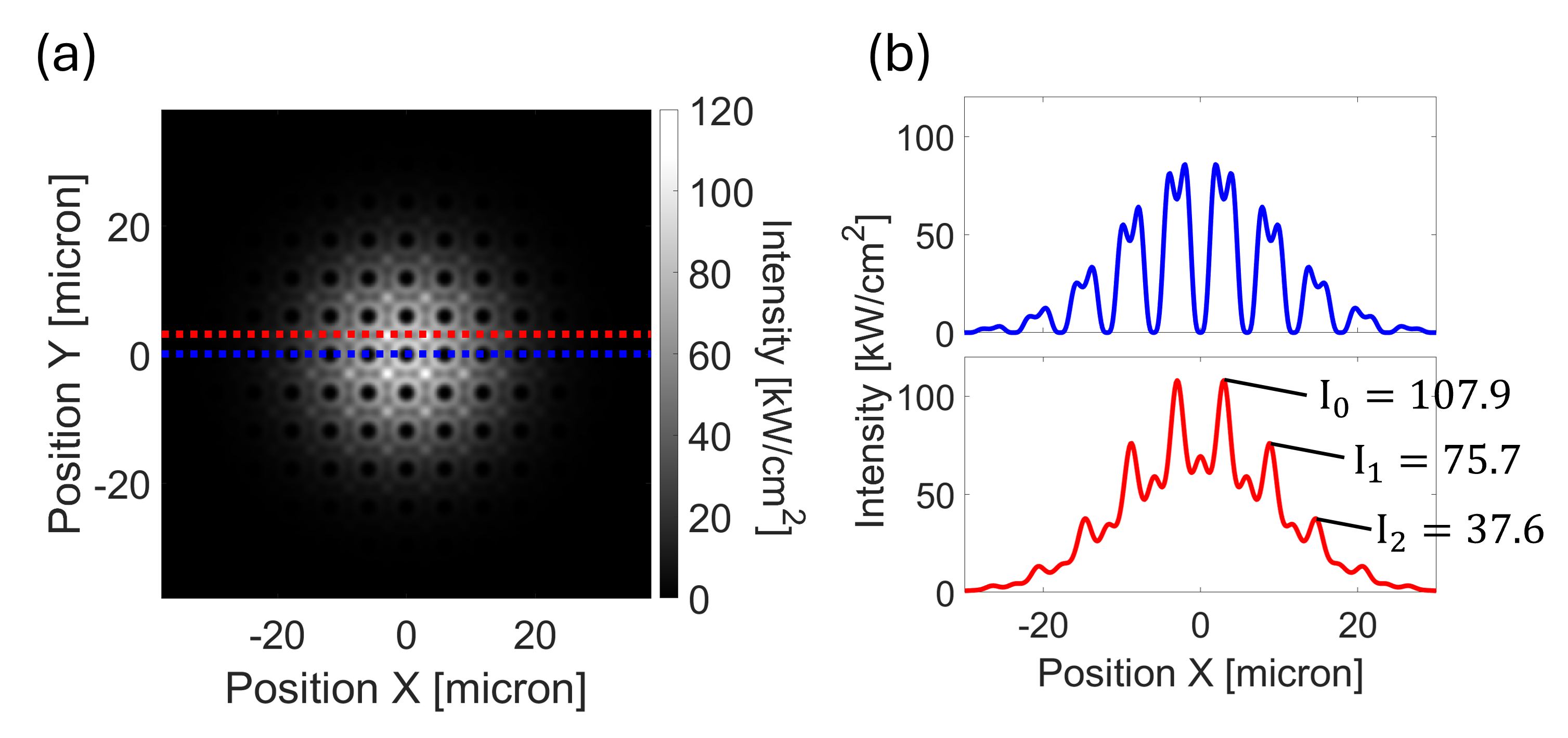} 
    \caption{\textbf{(a)} The calculated intensity profile at the image plane with a Gaussian beam input on the mask. The total power at the image plane is around 450 mW. \textbf{(b)} The calculated intensity line profiles of the dark traps (blue) and the bright traps (red).}
    \label{fig:intensity profile with gaussian input}
\end{figure}

In Fig. \ref{fig:intensity profile with gaussian input}(a), we calculated the intensity profile in the image plane, where total power is around 450 mW. Fig. \ref{fig:intensity profile with gaussian input}(b) shows the line profiles through the array for dark traps (blue) and bright traps (red). For the Rb atoms in the bright traps, the intensities of the trap centers change substantially over the array. For the Cs atoms in the dark traps, the intensities at the trapping sites are all close to zero. The light shift is proportional to the laser intensity at the trapping site, so the light shift of the Rb array is less homogeneous than the Cs array when Gaussian beam is used on the intensity mask. A top-hat beam generator can greatly improve the homogeneity of the bright trap array.

\end{document}